\documentclass[10pt,twocolumn,letterpaper]{article}

\usepackage[camera ready]{cvpr}      

\usepackage{graphicx}
\usepackage{amsmath}
\usepackage{amssymb}
\usepackage{booktabs}
\usepackage{subcaption}
\usepackage{makecell}
\usepackage{multirow}

\usepackage[pagebackref,breaklinks,colorlinks]{hyperref}

\usepackage[capitalize]{cleveref}
\crefname{section}{Sec.}{Secs.}
\Crefname{section}{Section}{Sections}
\Crefname{table}{Table}{Tables}
\crefname{table}{Tab.}{Tabs.}

\def\cvprPaperID{*****} 
\def\confName{CVPR}
\def\confYear{2024}

\begin{document}

\title{BeatDance: A Beat-Based Model-Agnostic Contrastive Learning Framework for Music-Dance Retrieval}

\author{
Kaixing Yang\\
Renmin University of China\\
{\tt\small yangkaixing@ruc.edu.cn}
\and
Xukun Zhou\\
Renmin University of China\\
{\tt\small xukun\textunderscore zhou@ruc.edu.cn}
\and
Xulong Tang \\
The University of Texas at Dallas \\
{\tt\small Xulong.Tang@UTDallas.edu}
\and
Ran Diao\\
Renmin University of China\\
{\tt\small diaoran@ruc.edu.cn}
\and
Hongyan Liu \\
Tsinghua University \\
{\tt\small liuhy@sem.tsinghua.edu.cn}
\and
Jun He\thanks{Corresponding authors}\\
Renmin University of China\\
{\tt\small hejun@ruc.edu.cn}
\and
Zhaoxin Fan\textsuperscript{$\ast$}\\
Renmin University of China\\
Psyche AI Inc\\
{\tt\small fanzhaoxin@ruc.edu.cn}
}

\maketitle

\begin{abstract}
Dance and music are closely related forms of expression, with mutual retrieval between dance videos and music being a fundamental task in various fields like education, art, and sports. However, existing methods often suffer from unnatural generation effects or fail to fully explore the correlation between music and dance. To overcome these challenges, we propose BeatDance, a novel beat-based model-agnostic contrastive learning framework. BeatDance incorporates a Beat-Aware Music-Dance InfoExtractor, a Trans-Temporal Beat Blender, and a Beat-Enhanced Hubness Reducer to improve dance-music retrieval performance by utilizing the alignment between music beats and dance movements. We also introduce the Music-Dance (MD) dataset, a large-scale collection of over 10,000 music-dance video pairs for training and testing. Experimental results on the MD dataset demonstrate the superiority of our method over existing baselines, achieving state-of-the-art performance. The code and dataset will be made public available upon acceptance.
\end{abstract}
\begin{figure}[t]
  \centering
  \includegraphics[width=0.49\textwidth]{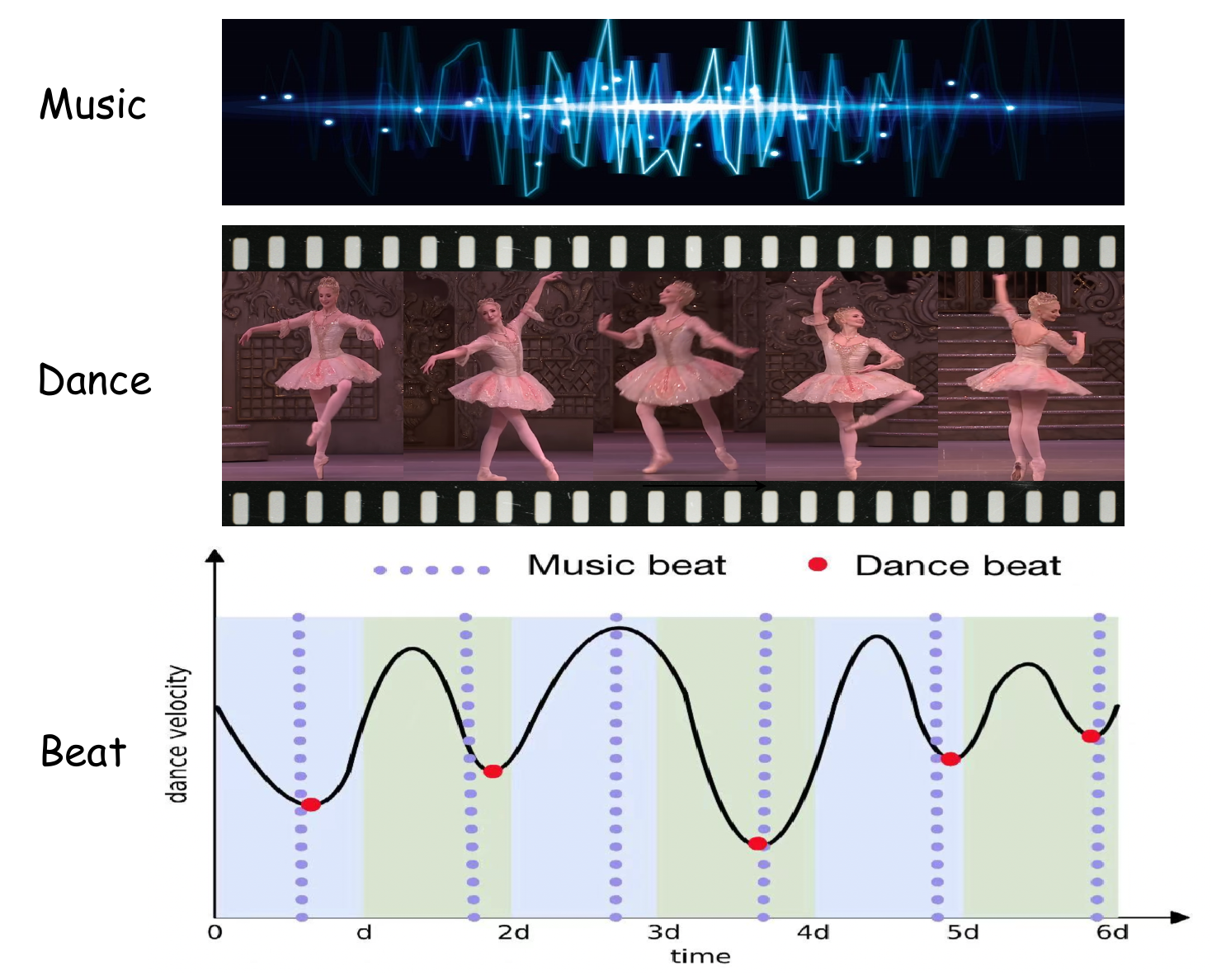}
   \vspace{-0.25in}
  \caption{This figure represents music, dance, and beat visualization from top to bottom. Red dots indicate occurrence of dance beats, while purple vertical lines represent occurrence of music beats. It is evident that there exists a certain degree of correspondence between dance beats and music beats.}
  \label{fig:intro}
  \vspace{-0.25in}
\end{figure}

\section{INTRODUCTION}
\par
Dance, as a significant art form, not only embodies human beauty and emotion but also serves as a crucial medium for cultural inheritance and communication. In recent years, with the rapid advancement of the Internet, the availability and impact of dance videos have witnessed a remarkable surge, providing audiences with diverse and captivating dance experiences. Consequently, the demand for large scale music-dance retrieval has grown exponentially, holding immense practical value for dance practitioners, encompassing areas such as dance education, art creation, and sports training.
\par
Existing approaches for obtaining music/dance from dance/music can be broadly categorized into generation-based and retrieval-based methods. While generation-based methods\cite{kim2022brand, li2021ai, valle2021transflower, siyao2022bailando} have shown significant progress in recent years, they encounter certain inherent challenges such as unnatural generation effects and limitations in generating diverse data types. For instance, in mainstream Music2Dance methods, only human key points are generated, neglecting factors like background and clothing. Similarly, in Dance2Music approaches\cite{vanwavenet, engel2018gansynth, goel2022s}, models with better performance often generate MIDI scores, overlooking the richness of human voice, background sound, and other audio details. On the other hand, retrieval-based methods naturally address these issues. Although music-dance retrieval has received comparatively less attention, there have been notable advancements\cite{tsuchida2019query, yu2022self}, but not fully explored correlation between dance and music.
\par
Generally, dancers synchronize their body movements with the rhythm of music, expressing their emotions and offering audiences a rich artistic experience, where the "beat" in dance and music serves as the most important information, as illustrated in Figure \ref{fig:intro}. Motivated by this observation, we propose BeatDance, a novel beat-based model-agnostic contrastive learning framework. In BeatDance, the concept of beat alignment between music and dance is fully utilized to enhance the model's focus on individuals. By incorporating temporal human pose information, representing the music beat, the model becomes more attuned to capturing the nuances of dancers' movements and allows for a stronger connection between the rhythmic elements of dance and music. Hence, it the retrieval performance could be significantly increased.

\par
BeatDance comprises three key blocks: the Beat-Aware Music-Dance InfoExtractor, the Trans-Temporal Beat Blender, and the Beat-Enhanced Hubness Reducer. In the InfoExtractor block, pre-trained models and methods are employed to extract rich information, including global features(CLIP\cite{radford2021learning}/MERT\cite{li2023mert}), music beat, and dance beat. The Feature Alignment module is utilized to unify the dimensions of these extracted features. The Beat Blender block involves sending the features to their respective Trans-Temporal Process modules to obtain trans-temporal features. These trans-temporal features are then blended with the global features using the Beat-Enhanced Feature Fusion module, and beat-guided features are obtained through the Beat-Guided Information Extraction module. To address the Hubness problem, the Beat-Enhanced Hubness Reducer block employs a query bank to normalize the similarity matrix during the inference phase, thereby alleviating issues associated with hubness. Additionally, we introduce the Music-Dance dataset(MD dataset), the first large-scale dataset specifically designed for the music-dance retrieval task. This dataset is sourced from Bilibili\cite{bilibili}, a popular video-sharing platform in China, covering the period from May 2018 to September 2023. It comprises 12,000 curated dance-music pairs with over 100,000 likes, encompassing various dance and music genres. Experimental results on the MD dataset demonstrate the superiority of our method compared to existing baselines, achieving state-of-the-art performance. 
\par
Our main contributions are summarized as below:
\begin{itemize}
  \item We introduce BeatDance, a novel beat-based model-agnostic contrastive learning framework that effectively utilizes the beat alignment information between music and dance to enhance the music-dance retrieval task.
  \item To facilitate the learning of music-dance correlation, BeatDance incorporates the Beat-Aware Music-Dance InfoExtractor, the Trans-Temporal Beat Blender, and the Beat-Enhanced Hubness Reducer. These modules work synergistically to jointly capture and leverage the relationship between music and dance.
  \item To evaluate and benchmark existing methods, we present the MD dataset, the first large-scale music-dance retrieval dataset. This dataset encompasses a wide range of dance and music genres, providing a comprehensive evaluation platform. Experimental results on the MD dataset demonstrate the superior performance of our proposed method.
\end{itemize}

\section{RELATED WORK}
\subsubsection{Music2Dance}
Generating natural and realistic human motion from music is a challenging problem. In recent years, significant progress has been made in the field of music-to-dance motion generation using various neural network architectures such as CNNs\cite{zhuang2023gtn, sun2020deepdance, ye2020choreonet, zhuang2022music2dance}, RNNs\cite{alemi2017groovenet, huang2020dance, sun2020deepdance, tang2018dance}, GCNs\cite{ferreira2021learning, ren2020self}, GANs\cite{lee2019dancing, sun2020deepdance}, or Transformers\cite{kim2022brand, li2021ai, valle2021transflower, siyao2022bailando}. Typically, these music-to-dance methods are conditioned on multimodal inputs and generate the future sequence of human poses. However, these methods still face several challenges. First, they are limited to generating only human poses and do not consider other important factors in dance, such as costumes, backgrounds, and facial expressions. Second, the generated motions often suffer from issues such as discontinuity and teleportation. Third, research efforts have largely been focused on solo dances while overlooking multi-person dances, despite their significant importance in dance practice. Furthermore, with the abundance of internet data, direct retrieval dance from music yields excellent results while avoiding above issues. Therefore, this paper focuses on the research of music-dance retrieval.

\subsubsection{Dance2Music}
Generating melodious and harmonious music for a given video is a challenging task, and there are two main categories of methods to address this task: non-symbolic based and symbolic based. Non-symbolic methods generate audio directly in the waveform, which is the original form of audio\cite{vanwavenet, engel2018gansynth, goel2022s}. However, a second of audio waveform covers a significant amount of data due to its high frequency. Even utilizing intermediate audio representations\cite{vasquez2019melnet, kumar2019melgan, dhariwal2020jukebox}, it is still computationally expensive and prone to generate noise. Symbolic methods adopt a symbolic music modeling approach, such as 1D piano-roll\cite{dong2018musegan} and 2D event-based MIDI-like\cite{huang2018music} music representations, etc.\cite{muhamed2021symbolic, ren2020popmag}. However, harmonious resonance of different timbres of instruments is essential to produce beautiful music, but symbolic methods often simplify the timbre, resulting in relatively monotonous generated music. Moreover, given the wealth of available internet data, performing direct retrieval music from video leads to outstanding outcomes, circumventing above concerns. Consequently, our paper delves into the exploration of dance-music retrieval.

\begin{figure*}[!t]
  \centering
  \includegraphics[width=\textwidth]{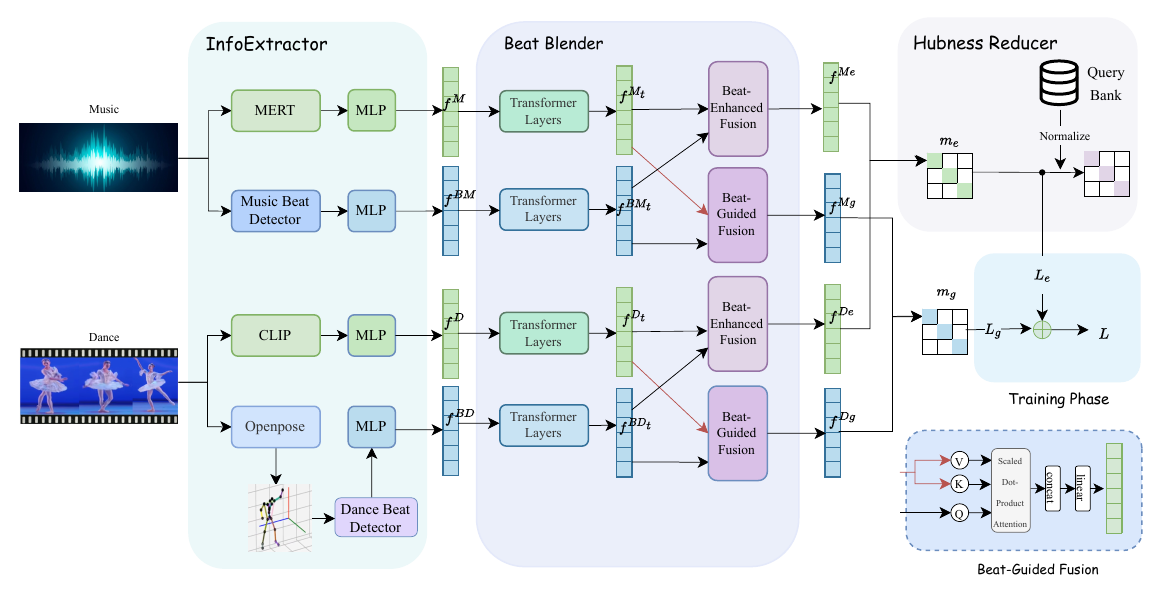}
   \vspace{-0.3in}
  \caption{Overview of BeatDance. We constructed a contrastive learning framework consisting of three blocks: InfoExtractor, Beat Blender, Beat-Enhanced Hubness Reducer. Specifically, given a music $m$ and a dance $d$, InfoExtractor first returns aligned global feature $f^{D}$ and $f^{M}$, beat feature of dance $f^{BD}$ and music $f^{BM}$. Then, Beat Blender processes them and returns beat-enhanced feature of music $f^{M_{e}}$ and dance $f^{D_{e}}$, beat-guided feature of music $f^{M_{g}}$ and dance $f^{D_{g}}$. Finally, we construct two similarity matrix $m_{e}$ and $m_{g}$ between two modality from beat-enhanced feature and beat-guided feature. In training phase, we utilize $m_{e}$ and $m_{g}$ to calculate beat-enhance loss $L_{e}$ and beat-guided loss $L_{g}$ for contrastive learning; in inference phase, we only send $m_{e}$, to Beat-Enhanced Hubness Reducer block and obtains normalized $m_{qbnorm}$, and computed retrieved sequences.}
  \label{fig:overview}
   \vspace{-0.2in}
\end{figure*}

\subsubsection{Music-Dance Retrieval}
Music-dance retrieval is a highly practical task in retrieval task, and music-dance retrieval can be considered as a subtask of video-music retrieval. In recent years, video-music retrieval have made significant progress\cite{cheng2023ssvmr, hong2018cbvmr, suris2022s, pandeya2021deep}. Typically, those above methods design a music and a video encoders to project raw modalities into a high-dimensional feature space, followed by contrastive learning training. However, video-music retrieval task primarily focus more on the high-level semantic consistency between the two modalities\cite{li2019query, mckee2023language}, while ignoring the real-time matching requirements between the two modalities. Relatively few to no researchers have paid attention to the field of Music-Dance retrieval\cite{tsuchida2019query, yu2022self}, and those who have mostly followed the traditional path of video-music retrieval, neglect strong beat correspondence between dance and music, and do not fully explore the correlation between music and dance. Moreover, we find there is no suitable large-scale dataset to benchmark music-dance retrieval methods. In this paper, we propose the BeatDance method and the MD dataset to solve the issue.

\section{METHODOLOGY}
\subsection{Overview}
\par
Our study involves two tasks: Music-Dance retrieval and Dance-Music retrieval, as Fig. \ref{fig:exp} shows. For Music-Dance retrieval task, we take a piece of music $m$ as input, and output the matching sequence of dance $\left\{ d_{1}, d_{2}... d_{n} \right\}$ from our database. For Dance-Music retrieval task, we take a piece of dance $d$ as input, and output the matching sequence of music $\left\{ m_{1}, m_{2}... m_{n} \right\}$ from our database.
\par
To better explore correlation between the music and dance modalities, we propose a Beat-Based Model-Agnostic contrastive learning framework called BeatDance, as Fig. \ref{fig:overview} shows. BeatDance consists of three blocks: Beat-Aware Music-Dance InfoExtractor, Trans-Temporal Beat Blender, and Beat-Enhanced Hubness Reducer. 
\par
In InfoExtractor block, we aim to acquire richer information and dimension unification. We send music $m$ and dance $d$ to it, and then obtain unified: music beat feature $f^{BM}$, dance beat feature $f^{BD}$, music global feature $f^{M}$, dance global feature $f^{D}$.
\begin{align}
\label{feature_alignment}
\begin{aligned}
    f^{D}, f^{BD} &= InfoExtractor_{d}(d) \\
    f^{M}, f^{BM} &= InfoExtractor_{m}(m)
\end{aligned}
\end{align}
\par
In Beat Blender block, we aim to leverage the strong correspondence between music beat and dance beat to better explore the correlation between Music and Dance. We send unified feature $f^{BM}, f^{BD}, f^{M}, f^{D}$ to it, and then get beat-enhanced feature $f_{M_{e}}, f_{D_{e}}$ and beat-guided feature $f_{M_{g}}, f_{D_{g}}$.
\begin{align}
\label{feature_alignment}
\begin{aligned}
    f^{D_{e}}, f^{D_{g}} &= BeatBlender_{d}(f^{D}, f^{BD}) \\
    f^{M_{e}}, f^{M_{g}} &= BeatBlender_{m}(f^{M}, f^{BM})
\end{aligned}
\end{align}
\par
In Hubness Reducer block, we aim to tackle the Hubness problem in retrieval task by constructing a query bank to normalize similarity matrix. Beat-Enhanced Hubness Reducer operates only during inference stage. We send our similarity matrix $m_{e}$ to it, and get a normalized matrix $m_{qbnorm}$:
\begin{equation}
\label{clip}
    m_{qbnorm} = Hubness Reducer(m_{e})
\end{equation}
\par
Finally, we can get ranked sequence by $m_{qbnorm}$ for music-to-dance or dance-to-music retrieval task.

\subsection{Beat-Aware Music-Dance InfoExtractor}
To tackle the challenge of music-dance retrieval, it is crucial to extract powerful features from both the dance video and the music, enabling the identification of their similarities. However, a naive approach would involve directly using global features extracted from CLIP\cite{radford2021learning} or MERT\cite{li2023mert} for retrieval purposes. While this approach seems straightforward, it has limitations. Pretrained CLIP\cite{radford2021learning} and MERT\cite{li2023mert} features are learned separately from other tasks and primarily focus on capturing global representations of images or music. Consequently, they may fail to capture the specific correlation between music and dance, hindering the effectiveness of music-dance retrieval. To overcome these limitations, we introduce the Beat-Aware Music-Dance InfoExtractor. 

\subsubsection{DanceInfo  Extractor}
\par
First, we calculate the CLIP\cite{radford2021learning} features for dance videos $d$. Then, we evenly divide CLIP\cite{radford2021learning} feature into $L$ intervals, and perform averaging operation on each interval. Finally, we obtain a dance feature $f^{d} \in R^{L \times d_{C}}$, which can represent entire dance, where $d_{C}$ represents dimension of CLIP feature. We denote process of obtaining CLIP feature as $\Gamma_{C}$:
\begin{equation}
\label{clip}
    f^{d} = \Gamma_{C}(d)
\end{equation}

\par
Second, we obtain human pose by sequence Openpose\cite{cao2017realtime}, then we calculate the dance beat $b^{d} \in R^{F_{d}}$ from pose sequence by Dance Beat Detector\cite{siyao2022bailando}, where $F_{d}$ represents frames number of dance video. To put it simple, the main idea of Dance Beat Detector is to consider the moments when the acceleration of movement is 0 as the beat points. We denote Dance Beat Detector as $\Phi_{d}$:
\begin{equation}
    b^{d} = \Phi_{d}(Openpose(d))
\end{equation}

\subsubsection{MusicInfo Extractor}
\par
First, we calculate the MERT\cite{li2023mert} features for music $m$. Then, we execute interval averaging operation as above CLIP features, and obtain a music feature $f^{m} \in R^{L \times d_{M}}$, which can represent entire music, where $d_{M}$ represents dimension of MERT\cite{li2023mert} feature, We denote process of obtaining CLIP\cite{radford2021learning} feature as $\Gamma_{M}$:
\begin{equation}
\label{mert}
    f^{m} = \Gamma_{M}(m)
\end{equation}
Second, we directly obtain the dance beat $b^{d} \in R^{F_{m}}$ by Music Beat Detector from Librosa\cite{mcfee2015librosa}, We denote Music Beat Detector as $\Phi_{m}$:
\begin{equation}
\label{music_beat}
    b^{m} = \Phi_{m}(m)
\end{equation}
\subsubsection{Feature Alignment}
Since the dimensions of $f^{d}$, $f^{m}$, $b^{d}$, and $b^{m}$ are all different, we need to implement a process of unification.
\par
With respect to beat, $b^m$ or $b^d$ can only take two possible values, 0 or 1, where 1 represents the presence of a beat and 0 represents its absence. Since beat is not a feature vector, segmenting and averaging as above methods would result in significant loss of information. To solve this problem, we first align the frame per second(fps) of $b^m$ and $b^d$, and then reshape them into $f^{bm}, f^{bd} \in R^{L \times d_{b}}$, respectively, where $d_{b}$ is dimension of beat feature. Additionally, We have processed all dance and music data to have equal durations, see Sec. \ref{dataset} for more details.
\par
Next, we use a two layers MLP to adjust their feature dimension of $f^{bm}, f^{bd}, f^{m}, f^{d}$, respectively, obtaining aligned features $f^{BM}, f^{BD}, f^{M}, f^{D} \in R^{L \times d_{u}}$, we denote this process as $\zeta$:
\begin{align}
\label{feature_alignment}
\begin{aligned}
    f^{D} &= \zeta_{D}(f^{d}) \\
    f^{M} &= \zeta_{M}(f^{m}) \\
    f^{BD} &= \zeta_{BD}(b^{d}) \\
    f^{BM} &= \zeta_{BM}(b^{m}) 
\end{aligned}
\end{align}

\subsection{Trans-Temporal Beat Blender}
As shown in Fig. \ref{fig:overview}, for both music and dance modalities, we extract two different kinds of features. However, simply concatenating or adding these features may not fully utilize their advantages. Moreover, it is important to consider capturing deep correlation between music and dance. To address these issues, we introduce a novel and efficient fusion block named Trans-Temporal Beat Blender.

\subsubsection{Trans-Temporal Processing}
Effective extraction of temporally spanning features significantly impacts the final results in both dance and music domains. In recent years, transformers have demonstrated remarkable success in extracting such features. Therefore, we employ four multi-layer transformer architecture to construct the Trans-Temporal Process module for $f^{D}, f^{M}, f^{BD}, f^{BM}$ respectively, and then obtain respective trans-temporal feature $f^{D_{t}}, f^{M_{t}}, f^{BD_{t}}, f^{BM_{t}} \in R^{L \times d_{u}}$, we denote this process as $\eta$.
\begin{align}
\label{TTFP}
\begin{aligned}
    f^{D_{t}} &= \eta_{D}(f^{D}) \\
    f^{M_{t}} &= \eta_{M}(f^{M}) \\
    f^{BD_{t}} &= \eta_{BD}(f^{BD}) \\
    f^{BM_{t}} &= \eta_{BM}(f^{BM}) 
\end{aligned}
\end{align}

\subsubsection{Beat-Enhanced Feature Fusion}
Due to the relatively weak correlation between music and dance features, it will introduce several challenges in retrieval tasks. However, it has been observed that music beat and dance beat exhibit a strong correspondence, indicating a potential avenue to resolve this problem. 
\par
To leverage this, a intuitive way is to use element-wise addition, but it fails to effectively capture cross-impact and non-linear relationships between features. Meanwhile, element-wise multiplication precisely addresses this issue\cite{he2017neural}, but is highly susceptible to noise interference. Thus, we combine above two method to achieve Beat-Enhanced Feature Fusion:
\begin{align}
\label{TTFP}
\begin{aligned}
    f^{D_{e}} &= MLP([f^{D_{t}} \oplus f^{BD_{t}}, f^{D_{t}} \otimes f^{BD_{t}}]) \\
    f^{M_{e}} &= MLP([f^{M_{t}} \oplus f^{BM_{t}}, f^{M_{t}} \otimes f^{BM_{t}}]) \\
\end{aligned}
\end{align}
where $f^{M_{e}},f^{D_{e}} \in R^{L \times d_{u}}$, and $MLP$ is used to rectify dimension.

\subsubsection{Beat-Guided Information Extraction}
On the one hand, after enhance the beat-related information in
$f^{D_{t}}$ and $f^{M_{t}}$ through Beat-Enhanced Feature Fusion, we next propose to guide the learning of $f^{D}t$ and $f^{M_{t}}$ towards the direction containing beat-related information, utilizing the Beat-Guided Information Extraction module.

\par
We utilize a Multi-Head Attention layer to construct Beat-Guided Information Extraction module. In this module, we can consider $f^{BM_{t}}$ and $f^{BD_{t}}$ information to be a subset of $f^{M_{t}}$ and $f^{D_{t}}$ information, to get beat-guided feature, we can construct Key and Value from $f^{M_{t}}, f^{D_{t}}$, and Query from $f^{BM_{t}}, f^{BD_{t}}$, as XPool\cite{gorti2022x}, we take dance part as example, and music part is similar:
\begin{align}
\label{TTFP}
\begin{aligned}
    Q_{b} &=\mathrm{LN}\left(f^{BD_{t}}\right) W_{Q} \\
    K_{d} &=\mathrm{LN}\left(f^{D_{t}}\right) W_{K} \\
    V_{d} &=\mathrm{LN}\left(f^{D_{t}}\right) W_{V}
\end{aligned}
\end{align}
\begin{equation}
head_{i}=\mathrm{softmax}\left(\frac{Q_{b} K_{d}^{T}}{\sqrt{D_{p}}}\right) V_{d}
\end{equation}

\begin{equation}
f^{D_{g}}= [head_{1}, \ldots, head_{h}] W_{O}
\end{equation}
where $\mathrm{LN}$ is a Layer Normalization layer, and $W_{Q}, W_{K}, W_{V}, W_{O}$ are projection matrices, and $h$ is head number, $f^{D_{g}}, f^{M_{g}} \in R^{L \times d_{u}}$.
\par

\begin{figure*}[!t]
  \centering
  \includegraphics[width=0.98\textwidth]{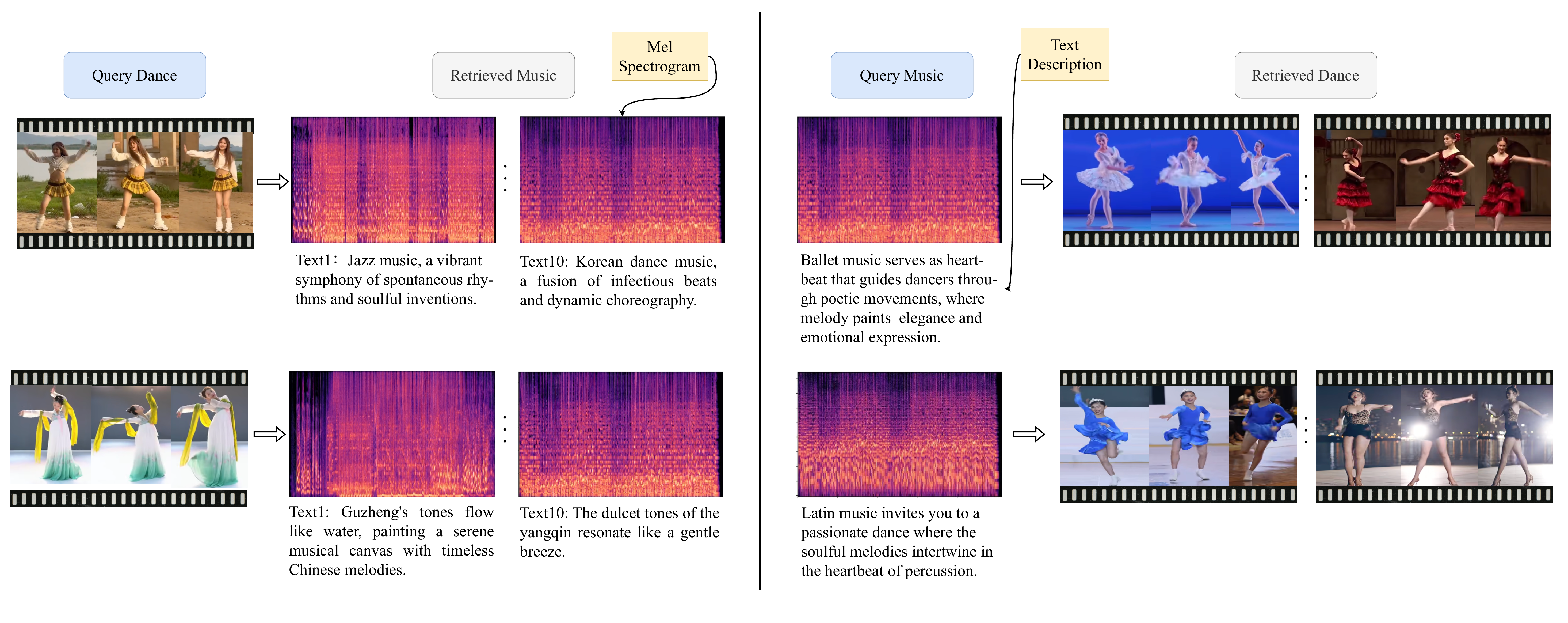}
   \vspace{-0.25in}
  \caption{BeatDance can effectively capture the underlying correspondence between dance and music. Given a piece of music/dance, the topk retrieved musics/dances exhibit a high degree of semantic similarity, such as in terms of dance/music style, emotional characteristics and etc.. It also demonstrates the strong expressive capability of BeatDance in feature extraction. Additionally, demonstration video of more experimental results can be found at \href{https://youtu.be/EsJAkDHDjgk}{YouTube-URL}.}
  \label{fig:exp}
  \vspace{-0.2in}
\end{figure*}

\subsection{Beat-Enhanced Hubness Reducer}
Despite the previous block's ability to effectively capture the correlation between music and dance, similar to other retrieval tasks\cite{bogolin2022cross}, a dance/music may always be reasonably matched to multiple music/dance, the Hubness Problem persists. Hubness problem refers to a phenomenon in which certain samples in high-dimensional data become central hubs, attracting a disproportionate number of nearest neighbors, which can lead to decreased retrieval accuracy, biased results, and difficulties in generalization. To tackle this challenge, we design the Beat-Enhanced Hubness Reducer block based on QBNorm\cite{bogolin2022cross}. Additionally, Beat-Enhanced Hubness Reducer only executes during inference phase.

\par
Specifically, we take music-dance retrieval as example. First, we construct a QueryBank set $S_{QB}$ from music in training/validation/test set. Second, we compute querybank-test similarity matrix $m_{qb\_t} \in R^{N_{qb} \times N_{t}}$ by query bank $S_{QB}$ and test dances set $S_{T^{d}}$, where $N_{t}$ and $N_{qb}$ represent number of test set and query bank, and then take the intersection of all $m \in S_d{QB}$'s top 1 matching dance to construct the Hubness-affecting dance set $S_{H_{d}}$. Third, we compute test similarity matrix $m_{e} \in R^{N_{t} \times N_{t}}$ by test music set $S_{T^{m}}$ and test 
 dance set $S_{T^{d}}$, and then select all Hubness-affected music $m_{H}$, whose top 1 matching dance is in Hubness-affecting dance set $S_{H_{d}}$, to construct Hubness-affected music set $S_{H_{m}}$. Additionally, the constructions of all similarity matrix stem from corresponding Beat-Enhanced feature. Fourth, we update test similarity matrix $m_{e}$:
\begin{equation}
m_{e}(i,j)=\frac{\exp \left(\beta \cdot m_{e}(i, j)\right)}{\mathbf{1}^{T} \exp \left[\beta \cdot m_{qb \_t}(j)\right]} \quad \text { if } music_{i} \in S_{H_{m}}
\end{equation}
where $i$ and $j$ represent the index of $S_{T^{m}}$ and $S_{T^{d}}$.

\par
Finally, we rename new matrix as QBNorm similarity matrix $m_{qbnorm}$, and can calculate ranked dance for each music from it. Hubness Reducer for dance-music retrieval is operated similarly.

\subsection{Training and Inference}
\subsubsection{Training}
During training stage, we execute contrastive learning, which encourages positive pairs to have a high similarity value, while vice versa. $f^{D_{e}}, f^{M_{e}}$ from the Beat-Enhanced Feature Fusion are used for computing Beat-Enhanced similarity matrix $m_{e}$. Then, we obtain Beat-Enhanced Loss $\mathcal{L}_{e}$ from $m_{e}$ by infoNCE\cite{oord2018representation} loss, and we perform similar operation to obtain Beat-Guided Loss $\mathcal{L}_{g}$:
\begin{equation}
\mathcal{L}_{e}^{m \rightarrow d}=-\frac{1}{B} \sum_{i=1}^{B} \log \frac{e^{s\left(f^{M_{e}}_{i}, f^{D_{e}}_{i}\right) \cdot \lambda}}{\sum_{j=1}^{B} e^{s\left(f^{M_{e}}_{i}, f^{D_{e}}_{j}\right) \cdot \lambda}} 
\end{equation}
\begin{equation}
\mathcal{L}_{g}^{m \rightarrow d}=-\frac{1}{B} \sum_{i=1}^{B} \log \frac{e^{s\left(f^{M_{g}}_{i}, f^{D_{g}}_{i}\right) \cdot \lambda}}{\sum_{j=1}^{B} e^{s\left(f^{M_{g}}_{i}, f^{D_{g}}_{j}\right) \cdot \lambda}} 
\end{equation}
\begin{equation}
\mathcal{L}^{m \rightarrow d}=\mathcal{L}_{e}^{m \rightarrow d} + \beta \times \mathcal{L}_{g}^{m \rightarrow d}
\end{equation}
where $s(m,d)$ represents cosine similarity, $B$ is batch size, $\lambda$ is temperature parameter, $\beta$ is a weighted hyperparameter. $\mathcal{L}^{d \rightarrow m}$ is computed symmetrically, and $\mathcal{L}=\mathcal{L}^{m \rightarrow d}+\mathcal{L}^{d \rightarrow m}$ is used for training our model.
\subsubsection{Inference}
During inference stage, we only construct similarity matrix $m_{e}$ through $f^{D_{e}}$ and $f^{M_{e}}$. Because Beat-Guided Information Extraction is designed solely to guide $f^{D_{t}}, f^{M_{t}}$ towards the direction that contains $f^{BD_{t}}, f^{BM_{t}}$ information during training phase, thus unnecessary to consider its influence during inference phase. Then, we send $m_{e}$ to Beat-Enhanced Hubness Reducer to get a normalized matrix $m_{qbnorm}$. Finally, we can calculate a ranked sequence from $m_{qbnorm}$ for music-dance or dance-music retrieval task.

\begin{table*}[!t]
\caption{Comparisons with state-of-the-art results on M-D dataset for music-to-dance and dance-to-music retrieval. Compared models include: CBVMR\cite{hong2018cbvmr}, XPool\cite{gorti2022x}, SCFEM \cite{nakatsuka2023content}, MQVR \cite{wang2022multi}, MVPt\cite{suris2022s}.}\label{baseline}
\vspace{-0.1in}
\renewcommand\arraystretch{1.2}
\centering
\resizebox{0.98\textwidth}{!}{
\begin{tabular}{l|cc|cc}
\Xhline{2pt}
\hline
\multirow{2}*{Method} & \multicolumn{2}{c|}{Music $\Longrightarrow$ Dance} & \multicolumn{2}{c}{Dance $\Longrightarrow$ Music} \\
\cline{2-5}
~ & Recall@1/10/50/100$\uparrow$ & MeanR/MedianR$\downarrow$ & Recall@1/10/50/100$\uparrow$ & MeanR/MedianR$\downarrow$\\
\Xhline{1.5pt}
CBVMR & 0.83/6.35/20.71/30.61 &	245.5/333.91 & 1.24/6.11/20.79/31.02 & 236.5/333.64 \\
SCFEM & 0.99/7.76/23.10/35.81 & 196.0/306.05 & 0.91/8.25/23.27/35.31 & 192.0/305.65 \\
MQVR & 1.65/8.91/26.90/39.60 & 152.5/263.80 & 1.24/9.49/26.90/39.11 & 152.0/265.36 \\
MVPt & 1.57/8.25/26.24/38.78 & 162.5/258.15 & 1.23/9.46/27.81/39.42 & 166.0/254.81 \\
XPool & 1.57/9.41/27.72/41.50 & 148.0/248.79 & 1.49/8.83/28.55/41.58 & 148.0/253.80 \\
\hline
BeatDance & \textbf{2.48/13.12/32.26/44.06} & \textbf{128.0/239.81} & \textbf{2.97/13.04/32.34/44.55} & \textbf{127.0/238.77} \\
\Xhline{2pt}
\end{tabular}
}
\end{table*}

\section{EXPERIMENT}
\subsection{Dataset}
\label{dataset}
To evaluate and benchmark existing methods in the Music-Dance retrieval task, we introduce M-D dataset, which is the first large-scale open-source dataset for this task. Fig. \ref{fig:exp} illustrates some examples of this dataset. The dataset is sourced from Bilibili\cite{bilibili}, the most popular video-sharing platform among young people in China. To ensure the quality and popularity of the dance videos, we collect videos uploaded between May 2018 and September 2023 in the dance category with over 100,000 likes. This ensures the excellence and popularity of the dataset.

The Music-Dance dataset encompasses various types of dance videos, including dance performances, tutorials, and practices in daily life. Through meticulous manual selection, we curate approximately 12,000 high-quality dance performance videos. The dataset is randomly shuffled and split into training, validation, and test sets in an 8:1:1 ratio. Statistical analysis of the dataset reveals that it contains both single-person and group dance performances, covering a wide range of dance genres such as Ballet, Contemporary, Hip-hop, Jazz, Tap, Latin, and more. It also includes a diverse selection of music genres, including Pop, Rock, Hip-hop, Electronic, Jazz, and others. Moreover, in addition to the dance and music video data, we provide dance beats extracted by Openpose\cite{cao2017realtime} and music beats extracted by Librosa\cite{mcfee2015librosa}. These beats are uniformly sampled at 10 frames per second (fps) and represented as binary values (1 for presence of beat, 0 for absence of beat). To ensure consistency in the analysis and evaluation of beat-based approaches, we consider a consistent 10-second segment from the middle of each dance video in our task. This ensures that all videos in the dataset have the same duration, allowing us to attribute any performance improvements solely to the presence of beats, independent of duration information.

\subsection{Evaluation}
Similar to other multi-modal retrieval tasks, such as text-video retrieval\cite{wang2022multi, gorti2022x}, video-music retrieval\cite{suris2022s, hong2018cbvmr}, we introduce Recall@K (higher is better) and Mean/Median Rank (lower is better) as evaluation metrics. To explore whether our method fully utilizes beats, we also introduce BS@K\cite{siyao2022bailando}(averaged Beat Similarity between ground truth and top k query results). We take dance-music retrieval as example:
\begin{equation}\label{BS}
BS_{d \rightarrow  m}=\frac{1}{\left|B^{m}\right|} \sum_{t^{m} \in B^{m}} \exp \left\{-\frac{\min _{t^{d} \in B^{d}}\left\|t^{d}-t^{m}\right\|^{2}}{2 \sigma^{2}}\right\}
\end{equation}
where, $t^{m}$ represents the moment when music beats occur. Likewise, $BS_{m \rightarrow d}$ is defined symmetrically in music-dance retrieval.

\subsection{Implementation Detail}
In our experiments, we employ CLIP's ViT-B/32\cite{radford2021learning} image encoder and MERT-95M\cite{li2023mert} as the base feature extractors. We initialize all encoder parameters using their pre-trained weights. The base features from CLIP\cite{radford2021learning} and MERT\cite{li2023mert} are precomputed, and the interval $L$ between base features is set to 10. The beat dim $d_{b}$ is set to 10. The feature unification dimension size is set to $d_u$=256. We initialize our logit scaling parameter $\lambda$ using the value from the pre-trained CLIP\cite{radford2021learning} model. For all transformers, we use a hidden dimension of 256, 6 layers, 4 heads, and a dropout rate of 0.3 (except for Beat-Guided Information Extraction, which uses a dropout rate of 0.3). During training, we set the batch size to 32 and the learning rate for the model parameters to 1e-5. We optimize our model for 150 epochs using the AdamW optimizer with a weight decay of 0.2. The learning rate is decayed using a cosine schedule. We use training set to construct query bank. Loss weight $\beta$ is set 0.4 for constrastive learning.

\subsection{Comparison}
To evaluate the performance of our method, we compared it with recent related works. Due to the limited availability of open-source code for video-music retrieval, let alone music-dance retrieval, we only reproduced the classic algorithms MVPt\cite{suris2022s} and CBVMR\cite{hong2018cbvmr} in this field. Additionally, we migrated models from other multimodal retrieval fields, such as XPool\cite{gorti2022x} and MQVR\cite{wang2022multi} in text-video retrieval and SCFEM\cite{nakatsuka2023content} in image-music retrieval. Specifically, for MVPt, since the music encoder(DeepSim) used in MVPt\cite{suris2022s} is not open-sourced, we replaced it with MERT\cite{li2023mert}. For CBVMR, due to the age of CBVMR, we replace its video encoder and music encoder with CLIP\cite{radford2021learning} and MERT\cite{li2023mert} respectively to ensure fairness. For XPool, we use averaged MERT\cite{li2023mert} feature of music instead of the CLIP feature of text. For MQVR, we first obtain MERT\cite{li2023mert}/CLIP\cite{radford2021learning} feature, and then uniformly divide it into 5 intervals, Averaged MERT\cite{li2023mert}/CLIP\cite{radford2021learning} feature of each interval represent one query in multi-query scene in MQVR. For SCFEM, we average the feature obtained by CLIP\cite{radford2021learning} over the time dimension to replace the original image feature.
\par

As shown in Tab. \ref{baseline}, our proposed BeatDance obviously outperforms all baseline methods, including CBVMR, SCFEM, MQVR, MVPt, and XPool, by a significant margin across various evaluation metrics.  Specifically, in the Music-to-Dance task, BeatDance achieves superior performance compared to other models, with Recall@1/10/50/100 values of 2.48/13.12/32.26/44.06, respectively. Additionally, BeatDance obtains lower MeanR/MedianR values, specifically 128.0/239.81. These results indicate that BeatDance significantly improves the accuracy of retrieving dance videos given music inputs. Similarly, in the Dance-to-Music task, BeatDance continues to outperform the baseline models. It achieves a Recall@1/10/50/100 of 2.97/13.04/32.34/44.55, surpassing all other models. The MeanR/MedianR values for BeatDance in this task are 127.0/238.77, which are lower compared to the baseline models. The superior performance of BeatDance can be attributed to its ability to capture and learn the correlation between music and dance videos more effectively. By considering beat alignment, BeatDance leverages the temporal structure and rhythmic patterns present in both the music and dance modalities. This allows the model to better align and synchronize the representations of music and dance, resulting in improved retrieval performance. The significant improvements achieved by BeatDance across all evaluation metrics establish its superiority over existing methods and position it as the current state-of-the-art (SOTA) approach in the field of music-dance retrieval. 

\begin{table*}[!t]
\caption{Effect of each component of BeatDance on M-D datasets for music-to-dance and dance-to-music retrieval.}
\vspace{-0.1in}
\renewcommand\arraystretch{1.2}
\label{ablation}
\centering
\resizebox{0.98\textwidth}{!}{
\begin{tabular}{l|cc|cc}
\Xhline{2.2pt}
\multirow{2}*{Method} & \multicolumn{2}{c|}{Music $\Longrightarrow$ Dance} & \multicolumn{2}{c}{Dance $\Longrightarrow$ Music} \\
\cline{2-5}
~ & Recall@1/10/50/100$\uparrow$ & MeanR/MedianR$\downarrow$ & Recall@1/10/50/100$\uparrow$ & MeanR/MedianR$\downarrow$\\
\Xhline{1.8pt}
Baseline & 2.15/11.87/29.29/42.33 & 142.0/256.28 & 2.48/12.05/28.38/41.50 & 145.0/257.73 \\
w/o Trans-Temporal Processing & \textbf{2.56}/11.80/27.81/40.43 & 166.0/281.25 & 2.56/11.88/27.48/39.93 & 163.5/284.41 \\
w/o Beat-Enhanced Feature Fusion & 1.98/12.21/29.13/42.41 & 140.0/258.32 & 2.39/11.37/29.62/41.34 & 147.0/260.39 \\
w/o Beat-Guided Information Extraction & 2.15/10.23/28.30/41.91 & 149.5/252.73 & 2.23/10.81/27.48/41.50 & 147.0/253.71 \\
w/o Hubness Reducer & 2.48/12.29/32.01/43.89 & 136.5/240.21 & 2.48/12.71/30.86/44.31 & 130.0/239.58 \\
Openpose\textrightarrow Mediapipe & 2.15/12.05/30.61/43.47 & 135.0/239.94 & 2.89/11.72/28.55/43.14 & 134.0/238.82 \\
\hline
Full BeatDance & 2.48\textbf{/13.12/32.26/44.06} & \textbf{128.0/239.81} &	\textbf{2.97/13.04/32.34/44.55} & \textbf{127.0/238.77} \\
\Xhline{2.2pt}
\end{tabular}
}
\end{table*}

\begin{table}[!t]
\caption{Effect of Fusion Mode in music-to-dance retrieval.}
\vspace{-0.1in}
\label{fusion}
\renewcommand\arraystretch{1.2}
\centering
\resizebox{0.49\textwidth}{!}{
\begin{tabular}{l|cc}
\Xhline{2.2pt}
Method & Recall@1/10/50/100$\uparrow$ & MeanR/MedianR$\downarrow$ \\
\Xhline{1.8pt}
Baseline & 2.15/11.87/29.29/42.33 & 142.0/256.28 \\
Beat Loss & 1.98/12.21/29.13/42.41 & 140.0/258.32 \\
Beat-Enhanced Feature Fusion(B) & 1.65/9.82/25.91/40.35 & 155.5/258.76 \\
Beat-Enhanced Feature Fusion(A) & 2.15/12.21/30.12/43.56 & 139.5/245.26 \\
Beat-Guided Information Extraction & 2.15/10.23/28.30/41.91 & 149.5/252.73 \\
\hline
BeatDance & \textbf{2.48/12.29/32.01/43.89} & \textbf{136.5/240.2} \\
\Xhline{2.2pt}
\end{tabular}
}
\end{table}

\begin{table}[!t]
\caption{Exploring BeatDance effect in music-to-dance retrieval. "+" means introduction of BeatDance.}
\vspace{-0.1in}
\renewcommand\arraystretch{1.2}
\centering
\resizebox{0.49\textwidth}{!}{
\begin{tabular}{l|ccc}
\Xhline{2.2pt}
Method & Recall@1/10/50/100$\uparrow$ & MeanR/MedianR$\downarrow$ & BS@1/5 $\uparrow$\\
\Xhline{1.8pt}
CBVMR & 0.83/6.35/20.71/30.61 & 245.5/333.91 & 85.11/84.97 \\
CBVMR+ & \textbf{0.99/8.33/23.93/37.21} & \textbf{179.5/276.02} &  \textbf{85.32/85.13} \\
\hline
XPool & 1.57/9.41/27.72/41.50 & 148.0/248.79 & 85.11/85.00 \\
XPool+ & \textbf{2.15/10.40/29.21/42.57} & \textbf{140.5/239.08} & \textbf{85.26/85.04} \\
\hline
Baseline & 2.15/11.87/29.29/42.33 & 142.0/256.28 & 85.15/85.16 \\
Baseline+ & \textbf{2.56/11.88/31.60/44.22} & \textbf{129.0/234.11} & \textbf{85.30/85.18} \\
\Xhline{2.2pt}
\end{tabular}
}
\label{model_agnostic}
\end{table}

\subsection{Ablation Study}
\subsubsection{Trans-Temporal Processing}
To better capture the trans-temporal information in music and dance related feature, we propose Trans-Temporal Processing. As shown in Tab. \ref{ablation}, the introduction of it makes great improvement in Recall@1/10/50/100(+2.55 in average) and Median/Mean Rank(+40.40 in average), which demonstrates its great effectiveness.
\subsubsection{Beat-Enhanced Feature Fusion}
To better enhance global information with corresponding beat information, we propose Beat-Enhance Fusion. As shown in Tab. \ref{ablation}, the introduction of it makes great improvement in Recall@1/10/50/100(+1.80 in average) and Median/Mean Rank(+18.03 in average), which demonstrates its great effectiveness.
\subsubsection{Beat-Guided Information Extraction}
To better guided music and dance related feature training direction containing beat information, we propose Beat-Guided Information Extraction. As shown in Tab. \ref{ablation}, the introduction of it makes great improvement in Recall@1/10/50/100(+2.53 in average) and in Median/Mean Rank(+17.34 in average), which demonstrates its great effectiveness. 
\subsubsection{Beat-Enhanced Hubness Reducer}
To address the Hubness problem, we design Beat-Enhanced Hubness Reducer. As shown in Tab. \ref{ablation}, the introduction of it makes great improvement in Recall@1/10/50/100(+0.47 in average) and in Median/Mean Rank(+3.18 in average), which demonstrates its great effectiveness.
\subsubsection{Pose Estimatior}
In the process of generating Dance Beats, pose estimation plays an important role, and we explore two popular methods Openpose and Mediapipe as our Pose Estimators. From Tab. \ref{ablation}, it can be observed that the performance based on Openpose is improved in Recall@1/10/50/100(+1.28 in average) and in Median/Mean Rank(+3.55 in average), compared to Mediapipe. This is because our dataset includes both multi-person and single-person dances, and in multi-person dances, Mediapipe focuses only on one dancer, neglecting the influence of others.
\subsubsection{Fusion Mode}
It is well known that beat information is crucial in dance and music. How to effectively integrate beat information with related music and dance features is an important problem. Thus, we also explore other feature fusion methods in addition to BeatDance.In Tab. \ref{fusion}, Beat Loss represents the separate contrastive learning training of global features and beat features after passing through the Trans-Temporal Process module. Beat-Enhanced Process(B) denotes the process in which global features and beats are first processed through the Beat-Enhanced Feature Fusion module, followed by the Trans-Temporal Process module, and then subjected to contrastive learning training. Beat-Enhanced Feature Fusion (A) refers to the process where global features and beats are initially processed through their respective Trans-Temporal Process modules, followed by the Beat-Enhanced Feature Fusion module, and subsequently undergo contrastive learning training. Beat-Guided Information Extraction signifies the process in which global features and beats are processed through their respective Trans-Temporal Process modules, followed by the Beat-Guided Information Extraction module, before undergoing contrastive learning training. BeatDance represents the BeatDance without utilizing the Beat-Enhanced Hubness Reducer module.As Tab. \ref{fusion} shows, BeatDance significantly outperforms other fusion methods on all metrics and it can be observed that the standalone use of Beat-Guided Information Extraction and Beat-Enhanced Feature Fusion yields inferior results.

\subsection{Model Analysis}
\subsubsection{Beat Similarity Analysis}
BeatDance integrates beat information and global features, naturally enhancing the correspondence between dance and music at the Beat level. BS@K can be a effective metric for evaluating if beat information is effectively utilized. As Tab. \ref{model_agnostic} shows, the introduction of BeatDance resulted in a improvement in BS@K on all models. It is worth noting that the limited improvement can be attributed to two factors: the minor role of beat in the retrieval task and the inherent limitations of the computational formula\ref{BS}. Even when provided with an beat array consisting entirely of ones, averaged Beat Similarity between it and all ground truth can still reach 69.86\%.

\begin{table}[t]
\caption{Comparson with others on classification task, including CBVMR\cite{hong2018cbvmr}, XPool\cite{gorti2022x}, SCFEM \cite{nakatsuka2023content}, MVPt+\cite{suris2022s}.}
\vspace{-0.1in}
\renewcommand\arraystretch{1.2}
\centering
\label{dowmstream}
\tiny
\resizebox{0.49\textwidth}{!}{
\begin{tabular}{l|ccc}
\Xhline{1pt}
Method & Genre & Instrument & Mood \\
\Xhline{1pt}
CBVMR & 50.58 & 64.60 & 61.14 \\
SCFEM & 53.88 & 69.22 & 61.22 \\
MVPt & 54.37 & 67.90 & 62.05 \\
XPool & 54.54 & 66.91 & 62.05 \\
\hline
BeatDance & \textbf{57.10} & \textbf{70.38} & \textbf{63.86} \\
\Xhline{1pt}
\end{tabular}
}
\end{table}

\begin{figure}[!t]
  \begin{subfigure}{0.23\textwidth}
    \includegraphics[width=\linewidth]{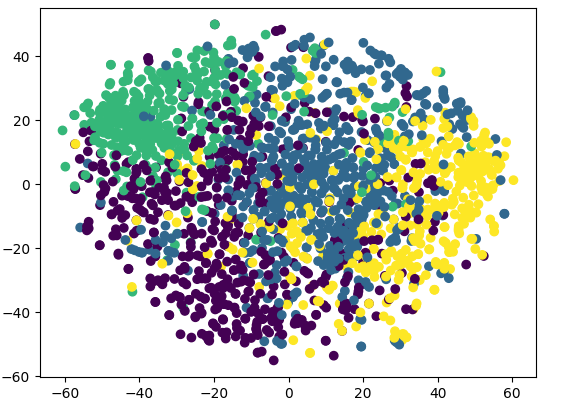}
    \caption{Music}
    \label{music_vis}
  \end{subfigure}
  \hfill
  \begin{subfigure}{0.23\textwidth}
    \includegraphics[width=\linewidth]{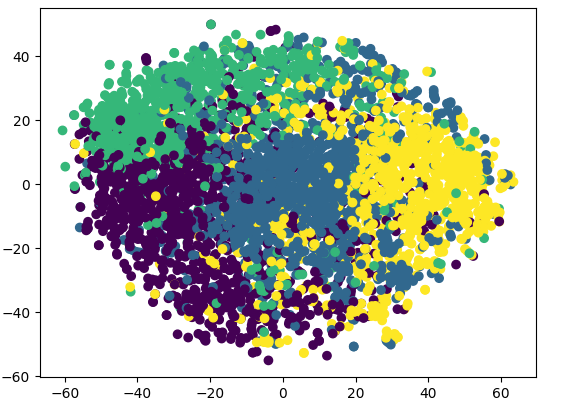}
    \caption{Dance}
    \label{video_vis}
  \end{subfigure}
   \vspace{-0.07in}
  \caption{t-SNE\cite{van2008visualizing} visulization of learned features. 2000 randomly sampled data pairs are chosen. It can be observed that music representations and dance representations exhibit a remarkably high degree of similarity in their distribution.}
    \vspace{-0.25in}
\end{figure}

\subsubsection{Model Agnositic Analysis}
\par
It is worth noting that BeatDance is essentially a framework with good generality, which is easy to extend to other models. Therefore, we conduct extra experiments on CBVMR and XPool. As shown in Tab. \ref{model_agnostic}, BeatDance greatly improved efficiency of all models, demonstrating its strong generalizability.

\subsubsection{Downstream Task Analysis}
To validate the expressive power of feature vectors generated by BeatDance, we introduce three classification tasks: music genre classification, music emotion classification, and music instrument classification.  We first employ well-known PANN\cite{kong2020panns} to assign genre, mood, and instrument labels to music for classification task. There are a total of 7 emotion categories, 23 genre categories, and 18 instrument categories. Then, we append two MLP layers to feature extracted by each model for subsequent classification. Accuracy was used as the evaluation metric. As shown in Tab. \ref{dowmstream}, BeatDance outperformed other models significantly in all three tasks, demonstrating its strong information extraction capabilities.

\subsubsection{Feature Distribution Analysis}
To explore feature representation capabilities of BeatDance, we randomly select 2000 instances from our dataset and obtain music representations and dance representations after processing them with BeatDance. Firstly, we apply K-Means clustering to assign cluster labels to all representations. Subsequently, we employ t-SNE\cite{van2008visualizing} for dimensionality reduction, projecting the high-dimensional features into a two-dimensional space. Finally, we visualize all 2000 data points. Fig. \ref{music_vis} and Fig. \ref{video_vis} illustrate the visualizations of the music representations and video representations respectively. Remarkably, it can be observed that the music representations and dance representations exhibit a high degree of similarity in their distributions, providing evidence for the efficient feature representation capabilities of BeatDance.

\section{CONCLUSION}
In this work, we have introduced BeatDance, a novel beat-based model-agnostic contrastive learning framework designed to better explore correlation between music and dance. In BeatDance, the Beat-Aware Music-Dance InfoExtractor, the Trans-Temporal Beat Blender, and the Beat-Enhanced Hubness Reducer are proposed to jointly facilitate the music-dance retrieval performance. To facilitate future research endeavors, we have also introduced the M-D dataset, the first large-scale open-source dataset specifically curated for the music-dance retrieval task. This dataset encompasses a diverse range of dance and music genres, providing a valuable resource for researchers in this field. Our experimental results have demonstrated the superiority of our proposed method compared to other baselines in the music-dance retrieval domain. We believe that this pioneering work will inspire and encourage more researchers and practitioners to explore and advance the capabilities of music-dance retrieval systems.

\bibliographystyle{ieee_fullname}
\bibliography{aaai22}

\begin{thebibliography}{10}\itemsep=-1pt

\bibitem{alemi2017groovenet}
Omid Alemi, Jules Fran{\c{c}}oise, and Philippe Pasquier.
\newblock Groovenet: Real-time music-driven dance movement generation using artificial neural networks.
\newblock {\em networks}, 8(17):26, 2017.

\bibitem{bilibili}
{Bilibili}.
\newblock Bilibili, 2023.
\newblock 2023.9.30.

\bibitem{bogolin2022cross}
Simion-Vlad Bogolin, Ioana Croitoru, Hailin Jin, Yang Liu, and Samuel Albanie.
\newblock Cross modal retrieval with querybank normalisation.
\newblock In {\em Proceedings of the IEEE/CVF Conference on Computer Vision and Pattern Recognition}, pages 5194--5205, 2022.

\bibitem{cao2017realtime}
Zhe Cao, Tomas Simon, Shih-En Wei, and Yaser Sheikh.
\newblock Realtime multi-person 2d pose estimation using part affinity fields.
\newblock In {\em Proceedings of the IEEE conference on computer vision and pattern recognition}, pages 7291--7299, 2017.

\bibitem{cheng2023ssvmr}
Xuxin Cheng, Zhihong Zhu, Hongxiang Li, Yaowei Li, and Yuexian Zou.
\newblock Ssvmr: Saliency-based self-training for video-music retrieval.
\newblock In {\em ICASSP 2023-2023 IEEE International Conference on Acoustics, Speech and Signal Processing (ICASSP)}, pages 1--5. IEEE, 2023.

\bibitem{dhariwal2020jukebox}
Prafulla Dhariwal, Heewoo Jun, Christine Payne, Jong~Wook Kim, Alec Radford, and Ilya Sutskever.
\newblock Jukebox: A generative model for music.
\newblock {\em arXiv preprint arXiv:2005.00341}, 2020.

\bibitem{dong2018musegan}
Hao-Wen Dong, Wen-Yi Hsiao, Li-Chia Yang, and Yi-Hsuan Yang.
\newblock Musegan: Multi-track sequential generative adversarial networks for symbolic music generation and accompaniment.
\newblock In {\em Proceedings of the AAAI Conference on Artificial Intelligence}, volume~32, 2018.

\bibitem{engel2018gansynth}
Jesse Engel, Kumar~Krishna Agrawal, Shuo Chen, Ishaan Gulrajani, Chris Donahue, and Adam Roberts.
\newblock Gansynth: Adversarial neural audio synthesis.
\newblock In {\em International Conference on Learning Representations}, 2018.

\bibitem{ferreira2021learning}
Joao~P Ferreira, Thiago~M Coutinho, Thiago~L Gomes, Jos{\'e}~F Neto, Rafael Azevedo, Renato Martins, and Erickson~R Nascimento.
\newblock Learning to dance: A graph convolutional adversarial network to generate realistic dance motions from audio.
\newblock {\em Computers \& Graphics}, 94:11--21, 2021.

\bibitem{goel2022s}
Karan Goel, Albert Gu, Chris Donahue, and Christopher R{\'e}.
\newblock It’s raw! audio generation with state-space models.
\newblock In {\em International Conference on Machine Learning}, pages 7616--7633. PMLR, 2022.

\bibitem{gorti2022x}
Satya~Krishna Gorti, No{\"e}l Vouitsis, Junwei Ma, Keyvan Golestan, Maksims Volkovs, Animesh Garg, and Guangwei Yu.
\newblock X-pool: Cross-modal language-video attention for text-video retrieval.
\newblock In {\em Proceedings of the IEEE/CVF conference on computer vision and pattern recognition}, pages 5006--5015, 2022.

\bibitem{he2017neural}
Xiangnan He, Lizi Liao, Hanwang Zhang, Liqiang Nie, Xia Hu, and Tat-Seng Chua.
\newblock Neural collaborative filtering.
\newblock In {\em Proceedings of the 26th international conference on world wide web}, pages 173--182, 2017.

\bibitem{hong2018cbvmr}
Sungeun Hong, Woobin Im, and Hyun~S Yang.
\newblock Cbvmr: content-based video-music retrieval using soft intra-modal structure constraint.
\newblock In {\em Proceedings of the 2018 ACM on international conference on multimedia retrieval}, pages 353--361, 2018.

\bibitem{huang2018music}
Cheng-Zhi~Anna Huang, Ashish Vaswani, Jakob Uszkoreit, Ian Simon, Curtis Hawthorne, Noam Shazeer, Andrew~M Dai, Matthew~D Hoffman, Monica Dinculescu, and Douglas Eck.
\newblock Music transformer: Generating music with long-term structure.
\newblock In {\em International Conference on Learning Representations}, 2018.

\bibitem{huang2020dance}
Ruozi Huang, Huang Hu, Wei Wu, Kei Sawada, Mi Zhang, and Daxin Jiang.
\newblock Dance revolution: Long-term dance generation with music via curriculum learning.
\newblock {\em arXiv preprint arXiv:2006.06119}, 2020.

\bibitem{kim2022brand}
Jinwoo Kim, Heeseok Oh, Seongjean Kim, Hoseok Tong, and Sanghoon Lee.
\newblock A brand new dance partner: Music-conditioned pluralistic dancing controlled by multiple dance genres.
\newblock In {\em Proceedings of the IEEE/CVF Conference on Computer Vision and Pattern Recognition}, pages 3490--3500, 2022.

\bibitem{kong2020panns}
Qiuqiang Kong, Yin Cao, Turab Iqbal, Yuxuan Wang, Wenwu Wang, and Mark~D Plumbley.
\newblock Panns: Large-scale pretrained audio neural networks for audio pattern recognition.
\newblock {\em IEEE/ACM Transactions on Audio, Speech, and Language Processing}, 28:2880--2894, 2020.

\bibitem{kumar2019melgan}
Kundan Kumar, Rithesh Kumar, Thibault De~Boissiere, Lucas Gestin, Wei~Zhen Teoh, Jose Sotelo, Alexandre De~Brebisson, Yoshua Bengio, and Aaron~C Courville.
\newblock Melgan: Generative adversarial networks for conditional waveform synthesis.
\newblock {\em Advances in neural information processing systems}, 32, 2019.

\bibitem{lee2019dancing}
Hsin-Ying Lee, Xiaodong Yang, Ming-Yu Liu, Ting-Chun Wang, Yu-Ding Lu, Ming-Hsuan Yang, and Jan Kautz.
\newblock Dancing to music.
\newblock {\em Advances in neural information processing systems}, 32, 2019.

\bibitem{li2019query}
Bochen Li and Aparna Kumar.
\newblock Query by video: Cross-modal music retrieval.
\newblock In {\em ISMIR}, pages 604--611, 2019.

\bibitem{li2021ai}
Ruilong Li, Shan Yang, David~A Ross, and Angjoo Kanazawa.
\newblock Ai choreographer: Music conditioned 3d dance generation with aist++.
\newblock In {\em Proceedings of the IEEE/CVF International Conference on Computer Vision}, pages 13401--13412, 2021.

\bibitem{li2023mert}
Yizhi Li, Ruibin Yuan, Ge Zhang, Yinghao Ma, Xingran Chen, Hanzhi Yin, Chenghua Lin, Anton Ragni, Emmanouil Benetos, Norbert Gyenge, et~al.
\newblock Mert: Acoustic music understanding model with large-scale self-supervised training.
\newblock {\em arXiv preprint arXiv:2306.00107}, 2023.

\bibitem{mcfee2015librosa}
Brian McFee, Colin Raffel, Dawen Liang, Daniel~P Ellis, Matt McVicar, Eric Battenberg, and Oriol Nieto.
\newblock librosa: Audio and music signal analysis in python.
\newblock In {\em Proceedings of the 14th python in science conference}, volume~8, pages 18--25, 2015.

\bibitem{mckee2023language}
Daniel McKee, Justin Salamon, Josef Sivic, and Bryan Russell.
\newblock Language-guided music recommendation for video via prompt analogies.
\newblock In {\em Proceedings of the IEEE/CVF Conference on Computer Vision and Pattern Recognition}, pages 14784--14793, 2023.

\bibitem{muhamed2021symbolic}
Aashiq Muhamed, Liang Li, Xingjian Shi, Suri Yaddanapudi, Wayne Chi, Dylan Jackson, Rahul Suresh, Zachary~C Lipton, and Alex~J Smola.
\newblock Symbolic music generation with transformer-gans.
\newblock In {\em Proceedings of the AAAI conference on artificial intelligence}, volume~35, pages 408--417, 2021.

\bibitem{nakatsuka2023content}
Takayuki Nakatsuka, Masahiro Hamasaki, and Masataka Goto.
\newblock Content-based music-image retrieval using self-and cross-modal feature embedding memory.
\newblock In {\em Proceedings of the IEEE/CVF Winter Conference on Applications of Computer Vision}, pages 2174--2184, 2023.

\bibitem{oord2018representation}
Aaron van~den Oord, Yazhe Li, and Oriol Vinyals.
\newblock Representation learning with contrastive predictive coding.
\newblock {\em arXiv preprint arXiv:1807.03748}, 2018.

\bibitem{pandeya2021deep}
Yagya~Raj Pandeya and Joonwhoan Lee.
\newblock Deep learning-based late fusion of multimodal information for emotion classification of music video.
\newblock {\em Multimedia Tools and Applications}, 80:2887--2905, 2021.

\bibitem{radford2021learning}
Alec Radford, Jong~Wook Kim, Chris Hallacy, Aditya Ramesh, Gabriel Goh, Sandhini Agarwal, Girish Sastry, Amanda Askell, Pamela Mishkin, Jack Clark, et~al.
\newblock Learning transferable visual models from natural language supervision.
\newblock In {\em International conference on machine learning}, pages 8748--8763. PMLR, 2021.

\bibitem{ren2020self}
Xuanchi Ren, Haoran Li, Zijian Huang, and Qifeng Chen.
\newblock Self-supervised dance video synthesis conditioned on music.
\newblock In {\em Proceedings of the 28th ACM International Conference on Multimedia}, pages 46--54, 2020.

\bibitem{ren2020popmag}
Yi Ren, Jinzheng He, Xu Tan, Tao Qin, Zhou Zhao, and Tie-Yan Liu.
\newblock Popmag: Pop music accompaniment generation.
\newblock In {\em Proceedings of the 28th ACM international conference on multimedia}, pages 1198--1206, 2020.

\bibitem{siyao2022bailando}
Li Siyao, Weijiang Yu, Tianpei Gu, Chunze Lin, Quan Wang, Chen Qian, Chen~Change Loy, and Ziwei Liu.
\newblock Bailando: 3d dance generation by actor-critic gpt with choreographic memory.
\newblock In {\em Proceedings of the IEEE/CVF Conference on Computer Vision and Pattern Recognition}, pages 11050--11059, 2022.

\bibitem{sun2020deepdance}
Guofei Sun, Yongkang Wong, Zhiyong Cheng, Mohan~S Kankanhalli, Weidong Geng, and Xiangdong Li.
\newblock Deepdance: music-to-dance motion choreography with adversarial learning.
\newblock {\em IEEE Transactions on Multimedia}, 23:497--509, 2020.

\bibitem{suris2022s}
D{\'\i}dac Sur{\'\i}s, Carl Vondrick, Bryan Russell, and Justin Salamon.
\newblock It's time for artistic correspondence in music and video.
\newblock In {\em Proceedings of the IEEE/CVF Conference on Computer Vision and Pattern Recognition}, pages 10564--10574, 2022.

\bibitem{tang2018dance}
Taoran Tang, Jia Jia, and Hanyang Mao.
\newblock Dance with melody: An lstm-autoencoder approach to music-oriented dance synthesis.
\newblock In {\em Proceedings of the 26th ACM international conference on Multimedia}, pages 1598--1606, 2018.

\bibitem{tsuchida2019query}
Shuhei Tsuchida, Satoru Fukayama, and Masataka Goto.
\newblock Query-by-dancing: a dance music retrieval system based on body-motion similarity.
\newblock In {\em MultiMedia Modeling: 25th International Conference, MMM 2019, Thessaloniki, Greece, January 8--11, 2019, Proceedings, Part I 25}, pages 251--263. Springer, 2019.

\bibitem{valle2021transflower}
Guillermo Valle-P{\'e}rez, Gustav~Eje Henter, Jonas Beskow, Andre Holzapfel, Pierre-Yves Oudeyer, and Simon Alexanderson.
\newblock Transflower: probabilistic autoregressive dance generation with multimodal attention.
\newblock {\em ACM Transactions on Graphics (TOG)}, 40(6):1--14, 2021.

\bibitem{vanwavenet}
A{\"a}ron van~den Oord, Sander Dieleman, Heiga Zen, Karen Simonyan, Oriol Vinyals, Alex Graves, Nal Kalchbrenner, Andrew Senior, and Koray Kavukcuoglu.
\newblock Wavenet: A generative model for raw audio.
\newblock In {\em 9th ISCA Speech Synthesis Workshop}, pages 125--125.

\bibitem{van2008visualizing}
Laurens Van~der Maaten and Geoffrey Hinton.
\newblock Visualizing data using t-sne.
\newblock {\em Journal of machine learning research}, 9(11), 2008.

\bibitem{vasquez2019melnet}
Sean Vasquez and Mike Lewis.
\newblock Melnet: A generative model for audio in the frequency domain.
\newblock {\em arXiv preprint arXiv:1906.01083}, 2019.

\bibitem{wang2022multi}
Zeyu Wang, Yu Wu, Karthik Narasimhan, and Olga Russakovsky.
\newblock Multi-query video retrieval.
\newblock In {\em European Conference on Computer Vision}, pages 233--249. Springer, 2022.

\bibitem{ye2020choreonet}
Zijie Ye, Haozhe Wu, Jia Jia, Yaohua Bu, Wei Chen, Fanbo Meng, and Yanfeng Wang.
\newblock Choreonet: Towards music to dance synthesis with choreographic action unit.
\newblock In {\em Proceedings of the 28th ACM International Conference on Multimedia}, pages 744--752, 2020.

\bibitem{yu2022self}
Jiashuo Yu, Junfu Pu, Ying Cheng, Rui Feng, and Ying Shan.
\newblock Self-supervised learning of music-dance representation through explicit-implicit rhythm synchronization.
\newblock {\em arXiv preprint arXiv:2207.03190}, 2022.

\bibitem{zhuang2023gtn}
Haolin Zhuang, Shun Lei, Long Xiao, Weiqin Li, Liyang Chen, Sicheng Yang, Zhiyong Wu, Shiyin Kang, and Helen Meng.
\newblock Gtn-bailando: Genre consistent long-term 3d dance generation based on pre-trained genre token network.
\newblock In {\em ICASSP 2023-2023 IEEE International Conference on Acoustics, Speech and Signal Processing (ICASSP)}, pages 1--5. IEEE, 2023.

\bibitem{zhuang2022music2dance}
Wenlin Zhuang, Congyi Wang, Jinxiang Chai, Yangang Wang, Ming Shao, and Siyu Xia.
\newblock Music2dance: Dancenet for music-driven dance generation.
\newblock {\em ACM Transactions on Multimedia Computing, Communications, and Applications (TOMM)}, 18(2):1--21, 2022.

\end{thebibliography}

\end{document}